\documentclass[trackchanges]{aastex7}

\usepackage{amsmath,amssymb, amsthm,amstext}
\usepackage{graphicx}
\usepackage{longtable}
\usepackage{tablefootnote}
\usepackage{caption}
\usepackage{color}
\usepackage{array, enumerate}
\usepackage{bm}
\usepackage{multirow}
\usepackage{gensymb}
\usepackage{braket}
\usepackage{txfonts}
\usepackage{booktabs}
\usepackage{afterpage}
\usepackage{subfigure}
\usepackage{rotating}

\newcommand{\old}[1]{{\color{red}{{}}}}

\renewcommand{\cite}{\citep}

\begin{document}

\title{Correlation between Ultrahigh-Energy Neutrino KM3-230213A and Gamma-Ray Bursts\footnote{ R.~Wang \& B.-Q.~Ma,  \href{https://doi.org/10.3847/1538-4357/ae5f77}{Astrophys.~J. 1003 (2026) 168}}}

\author[0009-0002-2076-1256]{Ruiqi Wang}
\affiliation{School of Physics, Peking University, Beijing 100871, China}
\email{ruiqiwang@stu.pku.edu.cn}
\author[0000-0002-5660-1070]{Bo-Qiang Ma}
\affiliation{School of Physics, Zhengzhou University, Zhengzhou 450001, China}
\affiliation{School of Physics, Peking University, Beijing 100871, China}
\email[show]{mabq@pku.edu.cn}
\correspondingauthor{Bo-Qiang Ma}



\begin{abstract}
 The KM3NeT Collaboration reported the detection of a neutrino, designated as KM3-230213A, with a reconstructed energy peaking at 220 PeV and equatorial coordinates (J2000) of RA=$94.3\degree$ and Dec=$-7.8\degree$. As the highest-energy neutrino event documented to date, its astrophysical origin remains unascertained. Prior preliminary investigations have probed potential associations between this neutrino event and gamma-ray bursts (GRBs), factoring in the possibility of Lorentz invariance violation (LV). In this study, we perform a comprehensive analysis to explore correlations between KM3-230213A and all viable GRBs. We explicitly account for the angular uncertainties intrinsic to both the neutrino event and the respective GRBs. Our analysis identifies a larger set of correlated GRBs. For each associated GRB, we compute the LV scale, integrating uncertainties from redshift measurements and neutrino energy determinations to enhance the robustness of our findings. 
\end{abstract}


\section*{}
On 13 February 2023 at 01:16:47 UTC, the KM3NeT Collaboration detected a neutrino, designated KM3-230213A, with a reconstructed energy of 220 PeV (best-fit value) and a 68\% confidence interval of 110 PeV to 790 PeV. Its equatorial coordinates (J2000) are RA=$94.3\degree$ and Dec=$-7.8\degree$ with containment radii: R(50\%)=$1.2\degree$ and R(90\%)=$2.2\degree$, respectively~\cite{KM3NeT:2025npi}. KM3-230213A is the highest-energy neutrino ever observed, while its astrophysical source has not been established yet. 
Beyond general efforts to associate this neutrino with potential astrophysical origins, specific investigations have explored its possible connection to the GRB\,090401B~\cite{Amelino-Camelia:2025lqn}. In our prior work, we performed a preliminary spatial analysis where we constrained the angular separation between candidate gamma-ray bursts (GRBs) and the neutrino to within $1\degree$, $3\degree$ or $5\degree$, and computed the Lorentz invariance violation (LV) scale for each proposed association~\cite{Wang:2025lgn}.

Lorentz invariance constitutes a foundational principle of special relativity, underpinning the fundamental symmetry of spacetime and ensuring the consistency of physical laws across inertial reference frames. LV refers to deviations from this symmetry, representing a potential departure from the standard framework of special relativity that has garnered significant interest in fundamental physics research. A key phenomenological signature of LV is the energy-dependent modification of the propagation velocities of ultra-high-energy particles, wherein their speeds may deviate from the universal speed of light ($c$) in a manner dependent on their energy. This energy-dependent velocity shift creates measurable effects in the arrival times and propagation characteristics of high-energy cosmic particles. Given that GRBs are known to emit high-energy photons with also possibile neutrinos over a broad energy spectrum, and these particles traverse cosmological distances before detection, they serve as exceptional astrophysical probes for testing LV effects. The simultaneous or correlated detection of high-energy neutrinos and photons from GRBs offers a unique opportunity to search for subtle LV-induced differences in their propagation speeds, making them ideal candidates for investigating such fundamental symmetry violations~\cite{Amelino-Camelia:1997ieq,Jacob:2006gn}.

Under LV, the dispersion relation for particles with energies significantly below the Planck scale (\( E \ll E_{\rm Pl} \approx 1.22 \times 10^{19} \, \rm{GeV} \)) undergoes modification, where the leading correction term arises from a Taylor series expansion. A general form of this modified dispersion relation is given by~\cite{Shao:2010wk}
\begin{equation}
E^2 \simeq p^2 c^2+m^2 c^4-s_n E^2\left(\frac{E}{E_{\mathrm{LV}, n}}\right)^n,
\end{equation}
where \( s_n = \pm 1 \) is a dimensionless sign factor determining the direction of the LV correction (positive for subluminal case and negative for superluminal case), \( n \) specifies the energy dependence of the LV term (typically \( n=1 \) or \( n=2 \) for linear or quadratic leading-order effects), and \( E_{\text{LV},n} \) represents the \( n \)-th order LV energy scale, which is to be constrained through experimental analysis. For ultra-high-energy neutrinos and photons, the mass term $ m^2 c^4 \ll p^2 c^2 $ and is also much less than the LV term, rendering it negligible. The modified propagation velocity $v(E)$, derived from $v=\partial E/\partial p$, is:
\begin{equation}
v(E)=c\left[1-s_n \frac{n+1}{2}\left(\frac{E}{E_{\mathrm{LV}, n}}\right)^n\right] .
\end{equation}
Such a speed variation can cause a propagation time difference between particles with different energies. Considering the cosmological expansion, the LV-induced time correction for two particles with energies $ E_h $ and $ E_l $ can be written as~\cite{Jacob:2008bw,Zhu:2022blp} 
\begin{equation}
\Delta t_{\rm LV}=s\cdot(1+z)\frac{K}{E_{\rm LV}},
\end{equation}
where $z$ is the redshift of the GRB source, $s=\pm1$ is the sign factor and
\begin{equation}
K=\frac{E_{\rm h}-E_{\rm l}}{H_0 }\frac{1}{1+z}\int^z_0\frac{(1+z^\prime) {\rm d} z^\prime}{\sqrt{\Omega_{\rm m}(1+z^\prime)^3+\Omega_\Lambda}},
\label{LV factor}
\end{equation}
is the LV factor for $n=1$. We adopt the cosmological parameters $[\Omega_{\rm m},\Omega_\Lambda]=[0.315^{+0.016}_{-0.017},0.685^{+0.017}_{-0.016}]$ and the Hubble expansion rate $H_0=67.3\pm 1.2~{\rm km\cdot s^{-1}\cdot Mpc^{-1}}$~\cite{ParticleDataGroup:2014cgo}.  
The observed arrival time difference $\Delta t_{\rm obs}$ between two particles depends on both LV correction time $\Delta t_{\rm LV}$ and the intrinsic time difference $\Delta t_{\rm in}$, the relation is~\cite{Ellis:2005sjy, Shao:2009bv}: 

\begin{equation}
\frac{\Delta t_{\rm obs}}{1+z}=\Delta  t_{\rm in}+s\cdot\frac{K}{E_{\rm LV}}.
\label{linear relation}
\end{equation} 

For neutrinos emitted in association with GRBs, the observed time difference $\Delta t_{\rm obs}$ is defined as the interval between the neutrino arrival time and the GRB trigger time, with the photon energy $E_l\approx 0$ (assuming photons travel at $c$). Under LV, the modified dispersion relation leads to a time delay. Eq.~(\ref{linear relation}) demonstrates a linear relationship between $\Delta t_{\rm obs}/(1+z)$ and $K$, if LV effects are present. 
The strategy is to correlate GRB neutrinos and photons by identifying coincident detections of both particles originating from the same GRB. Significant progress has been made in associating IceCube-detected neutrinos with GRBs to explore such correlations.  

Amelino-Camelia and collaborators analyzed IceCube ``shower" events within the energy range of 60–500 TeV, associating them with GRB candidates that occurred within a 3-day temporal window. Their studies revealed similar energy-dependent speed variation features between GRB neutrinos~\cite{Amelino-Camelia:2015nqa, Amelino-Camelia:2016fuh, Amelino-Camelia:2016ohi} and photons~\cite{Shao:2009bv, Zhang:2014wpb, Xu:2016zxi, Xu:2016zsa, Xu:2018ien, Liu:2018qrg}. Extending the temporal association window to 3 months, Huang and Ma~\cite{Huang:2018} identified four PeV-scale IceCube neutrinos linked to GRBs, with these events exhibiting consistency with energy-dependent speed variations. Such findings suggest signal for LV in cosmic neutrinos, with a corresponding LV energy scale of \( E_{\rm LV} \approx 6.4 \times 10^{17} \, \rm{GeV} \).  

Notably, both time-delayed and time-advanced events have been observed, which can be explained by distinct propagation properties between neutrinos and antineutrinos~\cite{Huang:2018}. This further implies potential CPT symmetry violation between neutrinos and antineutrinos, or an asymmetry between matter and antimatter~\cite{Zhang:2018otj}. Given the fundamental importance of these results, it is crucial to verify whether additional cosmic neutrino data continue to support the observed regularity~\cite{Amelino-Camelia:2016ohi, Huang:2018}. Subsequent studies reinforced this trend: Huang et al.~\cite{Huang:2019etr} found that 12 near-TeV IceCube track events align with the same energy-dependent pattern, and further work~\cite{Huang:2022xto} demonstrated that multi-TeV to PeV ``track" events could also be associated with GRBs under the same LV features observed in ``shower" events.  

A reanalysis~\cite{Amelino-Camelia:2022pja} of revised IceCube data~\cite{IceCube:2020wum} showed that for neutrinos subject to subluminal LV, the statistical significance of the results is even stronger than that reported in previous analyses~\cite{Amelino-Camelia:2016fuh, Amelino-Camelia:2016ohi, Huang:2018, Huang:2019etr, Huang:2022xto}. Most recently, the same research group associated the newly reported KM3-230213A event~\cite{KM3NeT:2025npi} with GRB\,090401B, which was observed 14 years prior~\cite{Amelino-Camelia:2025lqn}. This association constrains the subluminal LV energy scale to \( E_{\rm LV} = (3.97 \, \text{--} \, 9.60) \times 10^{17} \, \rm{GeV} \), consistent with the earlier result of \( E_{\rm LV} \sim 6.4 \times 10^{17} \, \rm{GeV} \). 

In a brief note~\cite{Wang:2025lgn} by our group, an attempt was made to associate this extra-ordinary neutrino KM3-230213A with potential GRBs within angular separation of $1\degree$, $3\degree$~ and $5\degree$~ respectively and the results suggest that a number of GRBs (\(\sim\)10) exhibits potential associations with KM3-230213A across a broad range of subluminal LV energy scales (\(E_{\rm LV} \sim (3\text{--}10) \times 10^{17}\,{\rm GeV}\)). The note also proposed the constraints $E_{\rm{LV}}\leq 5.3\times 10^{18}$~GeV for subluminal LV violation and $E_{\rm{LV}}\leq 5.6\times 10^{19}$~GeV for superluminal LV violation if KM3-230213A is a GRB neutrino. 

The purpose of the present study is to conduct a more rigorous analysis to investigate correlations between KM3-230213A and all plausible GRBs, explicitly accounting for the angular uncertainties inherent to both the neutrino event and the corresponding GRBs. 
Based on the above analysis, considering the extended time span and extra-high neutrino energy (220 PeV) of this study, subsequent calculations will neglect the intrinsic time difference $\Delta t_{\rm in}$ between  neutrino and GRB in Eq.~\ref{linear relation}, as its impact on the results is negligible.

Based on data from GRBweb \cite{coppin_grbweb_summary}, we perform a systematic analysis of GRB events detected by one or more satellite instruments (Fermi-GBM, Fermi-LAT, Swift-BAT\footnote{Swift-BAT refers to the initial burst trigger by the Burst Alert Telescope, while Swift-XRT refers to the subsequent X-ray afterglow detection by the X-Ray Telescope.}, Swift-XRT, MAXI, CALET, INTEGRAL, IPN, BeppoSAX, BATSE) within a temporal window spanning from +11615 days (27 April 1991) to -908 days (08 August 2025) relative to the neutrino arrival time (where positive values indicate times before the neutrino arrival and negative values indicate times after). This analytical framework comprehensively incorporates the positional uncertainties of both the neutrino and GRBs, enabling the identification of GRB events potentially associated with the neutrino. 

To screen candidate GRBs, we apply a spatial constraint using a two-dimensional circular Gaussian distribution~\cite{Amelino-Camelia:2016fuh}:  
\begin{equation}
P(\nu, \mathrm{GRB}) = \frac{1}{2 \pi \theta^2} \exp \left(-\frac{\Delta \phi^2}{2 \theta^2}\right)
\end{equation}  
where $\Delta \phi$ denotes the angular separation between the GRB and the neutrino. 
Positional uncertainties of GRBs are typically characterized by the 1$\sigma$ contour of a two-dimensional Gaussian density distribution, while neutrino positional uncertainties are commonly quantified using containment radii. To objectively and consistently combine these two different uncertainty conventions, we define two angular thresholds corresponding to different confidence levels.

For the primary criterion, we define 
$\theta = \sqrt{R_{50\%}^2 + \sigma_{\rm GRB}^2}$,  which combines the neutrino's 50\% containment radius ($R_{50\%} = 1.2^\circ$) with the GRB 1$\sigma$ positional error. 
For a two-dimensional Gaussian distribution, the 1$\sigma$ contour encloses 
39.3\% of the probability, closely matching the 50\% containment radius; GRBs satisfying $\Delta\phi < \theta$ thus fall within the combined 
$\sim$50\% confidence region (see Table~\ref{Table 1} and Figure 1).

For the extended criterion, we define $3\tilde{\theta} = \sqrt{R_{99\%}^2 + 9\sigma_{\rm GRB}^2}$, 
where $R_{99\%} = 3.0^\circ$ is the 99\% containment radius of KM3-230213A. 
This ensures that the extended search region corresponds to the combined $\sim$99\% confidence region, without exceeding the neutrino localization uncertainty when $\sigma_{\rm GRB}$ is small. 
GRBs satisfying $\Delta\phi < 3\tilde{\theta}$ are included as 
candidate associations at lower confidence (see Table~\ref{Table 2} and Fugure 2).

The $\theta$ sample represents the spatially most probable associations 
($\sim$50\% combined confidence), while the $3\tilde{\theta}$ sample serves as a completeness-oriented extension at the $\sim$99\% confidence level,  capturing potential associations with well-localized GRBs that may fall 
outside the narrower $\theta$ region.

Some GRBs in GRBweb~\cite{coppin_grbweb_summary} lack recorded positional errors; in such cases, we approximate the GRB positional error using the typical localization accuracy of the corresponding instrument~\cite{von_Kienlin_2020,Gehrels_2004,fermi_lat_overview,10.1117/12.409158,Rau:2005ge,Hurley_2013,Pal'shin_2013,Matsuoka_2009,Asaoka_2019}.

To quantify the plausibility of each neutrino--GRB spatial association, we calculate the chance coincidence probability
\begin{equation}
  P_{{\rm chance},i} \sim \frac{\Omega_i}{\Omega_{\rm sky}}, \quad
  \Omega_i = \pi\!\left(R_{50\%}^2 + \sigma_{{\rm GRB},i}^2\right),
  \label{eq:pchance}
\end{equation}
where $R_{50\%} = 1.2^\circ$ is the 50\% containment radius of KM3-230213A, $\sigma_{{\rm GRB},i}$ is the GRB localization error, and $\Omega_{\rm sky} = 41\,253\;\mathrm{deg}^2$ is the full-sky solid angle.
This quantity gives the fraction of the sky covered by the joint positional uncertainty region, and thus represents the probability that a randomly placed source would fall within this region by chance.

Table~\ref{tab:pchance} lists $P_{{\rm chance},i}$ for representative neutrino--GRB pairs spanning the full range of localization quality. Well-localized GRBs such as GRB\,090401B ($\sigma_{\rm GRB} \sim 0.00007^\circ$), GRB\,170610A, and GRB\,230402A yield $P_{{\rm chance},i} \sim 0.01\%$, suggesting that their spatial coincidence with KM3-230213A is unlikely to arise by chance. In contrast, poorly localized GRBs ($\sigma_{\rm GRB} \gtrsim 30^\circ$) have $P_{{\rm chance},i}$ up to $\sim 39\%$, and their apparent association should be interpreted with caution. Over the full $1\theta$ sample (54 pairs), $P_{{\rm chance},i}$ ranges from $\sim 0.09\%$ to $\sim 39\%$ with a median of $\sim 8\%$.

It is worth noting that GRB\,090401B, which has the smallest $P_{{\rm chance},i}$ in our sample, was independently identified by \citet{Amelino-Camelia:2025lqn} as a promising candidate for in-vacuo dispersion analysis with a combined $p$-value of 0.015.

\begin{table}[t]
\centering
\caption{Representative chance coincidence probabilities $P_{{\rm chance},i}$ for selected neutrino--GRB pairs.}
\label{tab:pchance}
\begin{tabular}{cccc}
\toprule
GRB Name & $\sigma_{\rm GRB}$ ($^\circ$) & $\Delta\phi$ ($^\circ$) & $P_{{\rm chance},i}$ \\
\midrule
GRB\,090401B  & 0.00007 & 1.41  & 0.011\% \\
GRB\,170610A  & 0.23     & 2.36  & 0.011\% \\
GRB\,230402A  & 0.40     & 2.35  & 0.012\% \\
GRB\,150213A  & 2.49     & 3.11  & 0.058\% \\
GRB\,940214C  & 3.17     & 2.78  & 0.087\% \\
GRB\,130213A  & 9.70     & 4.75  & 0.73\%  \\
GRB\,920711A  & 13.07    & 0.62  & 1.3\%   \\
GRB\,000704A  & 31.0     & 30.2  & 7.3\%   \\
GRB\,020212B  & 72.0     & 67.4  & 39\%    \\
\bottomrule
\end{tabular}
\end{table}

Since our search is purely spatial --- no temporal cut is applied beyond requiring the GRB to have occurred before the neutrino detection --- the discriminating power of $P_{{\rm chance},i}$ comes entirely from the angular localization precision.
This design choice reflects the nature of LV searches: the photon time delay can range from seconds to decades depending on the LV energy scale, making the time separation between a GRB and the neutrino an unreliable discriminant in our study.

During the detection process of neutrinos, both continuous and stochastic energy losses occur. The energy detected in track events represents only a fraction of the muon energy, which itself is only a portion of the neutrino total energy. Consequently, the detected energy is significantly lower than the actual neutrino energy. KM3NeT's reconstruction of neutrino energy accounts for these factors as well as potential errors during detection, providing a reconstructed neutrino energy with associated energy ranges at different confidence levels. We adopt the central value of the reconstructed neutrino energy, 220 PeV, to calculate the corresponding central value of $E_{\rm LV}$. Additionally, we use the energy range at the 68\% confidence level (110–790 PeV) to compute the corresponding $E_{\rm LV}$ range for each dataset.

For GRBs with unknown redshift \( z \), we adopt the average redshift values adopted in previous studies: \( z = 2.15 \) for ``long bursts" and \( z = 0.5 \) for ``short bursts". These values correspond to the median redshifts of the observed long and short GRB populations, respectively ~\cite{Kruhler:2012tz, Jakobsson:2005jc, annurev:/content/journals/10.1146/annurev-astro-081913-035926} and have been adopted in previous LV studies when spectroscopic redshifts are unavailable~\cite{Huang:2018,Huang:2022xto,Wang:2025lgn}. However, with the accumulation of more observational data, a subset of GRBs lacks not only redshift measurements but also duration information—key parameters for classifying long and short bursts. For such GRBs, we use \( z = 1.88 \) as the estimated average redshift, following the updated statistical result from recent work~\cite{10.3389/fspas.2023.1124317}.  

To account for the uncertainty in redshift estimates for these data points, all GRBs with unknown redshift are assigned an error range of \( 0.5z \) to \( 2z \) in this analysis. This range is chosen to effectively encompass the distinct redshift distributions of ``long bursts" and ``short bursts," ensuring that potential misclassification between the two classes does not introduce significant bias. We have also examined the impact of uncertainties in the cosmological parameters and find that their effect on the derived results is negligible compared to that induced by the redshift and the intrinsic energy. By incorporating both this redshift uncertainty and the intrinsic energy error reported for the GRB event KM3-230213A~\cite{KM3NeT:2025npi}, we aim to maximize the accuracy of the derived Lorentz violation parameter \( E_{\rm LV} \). We note that for GRB\,090401B, Amelino-Camelia et al.~\cite{Amelino-Camelia:2025lqn} adopted a photometric redshift of $z = 3.1$. In our analysis, since this photometric estimate has not been confirmed by spectroscopic observations, we apply a uniform treatment for all GRBs lacking spectroscopic redshift measurements to ensure consistency across the sample. Since $z = 3.1$ falls well within our assumed uncertainty range ($0.5z$ to $2z$, i.e., $z = 1.075$--$4.3$) for long bursts, the $E_{\rm LV}$ constraints derived for this burst already encompass the case of $z = 3.1$.

After accounting for the positional errors of the GRBs, the number of GRBs with an angular difference from KM3-230213A less than $\theta$ increases to 54, as shown in Table~2 and Figure~1. The constraint on the Lorentz-violating energy scale $E_{\rm LV}$, derived from the superluminal neutrino contribution associated with GRB\,240625B, has a central value of $5.4\times10^{18}~\rm GeV$. Taking into account the errors in energy and redshift, the corresponding range at the 68\% confidence level is $(14.0\text{--}358.9)\times10^{17}~\rm GeV$, indicating an upper limit of $E_{\rm LV}<3.59\times10^{19}~\rm GeV$ for the superluminal contribution ($\Delta\phi<\theta$). 
Similarly, the constraint on the Lorentz-violating energy scale $E_{\rm LV}$, derived from the subluminal neutrino contribution associated with GRB\,220618B*, has a central value of $2.7\times10^{18}~\rm GeV$. Accounting for uncertainties in energy and redshift, the corresponding range at the 68\% confidence level is $(6.4\text{--}199.6)\times10^{17}~\rm GeV$, indicating an upper limit of $E_{\rm LV}<1.996\times10^{19}~\rm GeV$ for the subluminal contribution ($\Delta\phi<\theta$). 

The constraint on $E_{\rm LV}$ is obtained by associating the KM3NeT neutrino event with a GRB at a certain time lag (absolute value). The upper limit of $E_{\rm LV}$ is obtained by associating the KM3NeT neutrino event with the GRB with smallest time lag, therefore the ``upper limit" should not be interpreted as a strict physical upper bount but rather than a best-case estimate depending on the assumed associations. As in our analysis, the time lag of the associated GRBs could be as large as possible (even to $\infty$, corresponding to $E_{\rm LV}\to 0$), so there is no lower limit of $E_{\rm LV}$.

Table~\ref{Table 2} and Figure 2 show the other 293 associated GRBs with angular offsets between $\theta$ and $3\tilde{\theta}$ from neutrino KM3-230213A. Under more relaxed restrictions, the constraint of $E_{\rm LV}$ for superluminal LV violation is from the association with GRB\,230402A. The central value is $E_{\rm{LV}}= 5.6\times 10^{19}~\rm GeV$ and the corresponding range with energy and redshift errors is  $(1.4\text{--}37.1)\times10^{19}~\rm GeV$, indicating an upper limit of $E_{\rm LV}<3.71\times10^{20}~\rm GeV$ for the superluminal contribution ($\Delta\phi<\theta$). 
Correspondingly, the constraint on the Lorentz-violating energy scale $E_{\rm LV}$ for subluminal neutrino contribution is given by GRB\,230126A. The central value is $E_{\rm{LV}}= 15.6\times 10^{19}~\rm GeV$, accounting for uncertainties in energy and redshift, the corresponding range at the 68\% confidence level is $(4.0\text{--}103.6)\times10^{19}~\rm GeV$, indicating an upper limit of $E_{\rm LV}<1.036\times10^{21}~\rm GeV$ for the subluminal contribution ($\Delta\phi<\theta$).

\begin{figure}[htbp]
    \centering
    \includegraphics[width=0.75\linewidth]{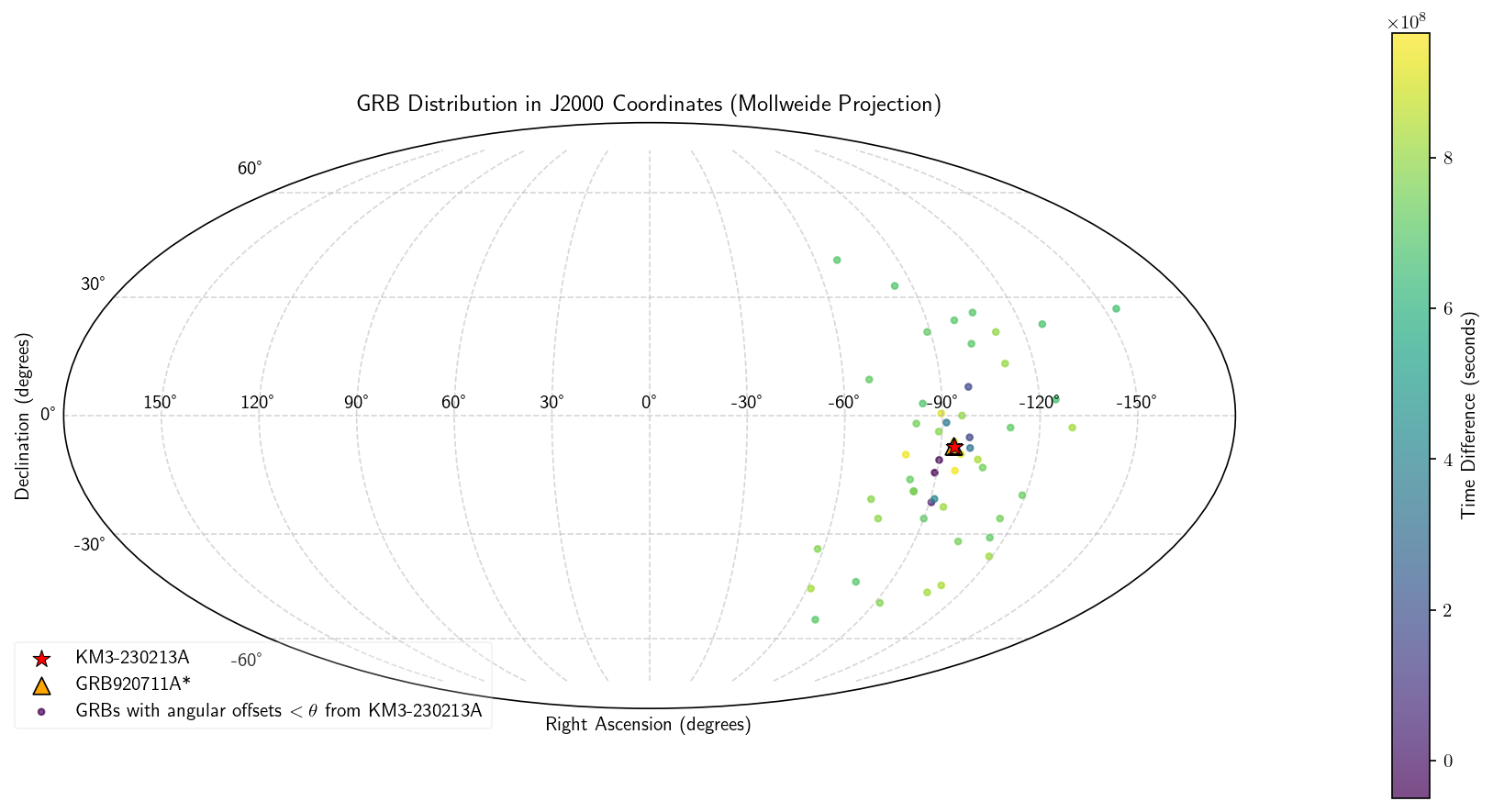}
    \caption{Mollweide projection showing the distribution of GRBs in J2000 coordinates, with colors representing time differences from KM3-230213A (marked by a red star). The orange triangle highlights GRB\,920711A*, the closest angular match in the $<\theta$ sample. All GRBs shown have angular offsets within $\theta$.
    }
    \label{fig:sky map1}
\end{figure}

\begin{figure}[htbp]
    \centering
    \includegraphics[width=0.75\linewidth]{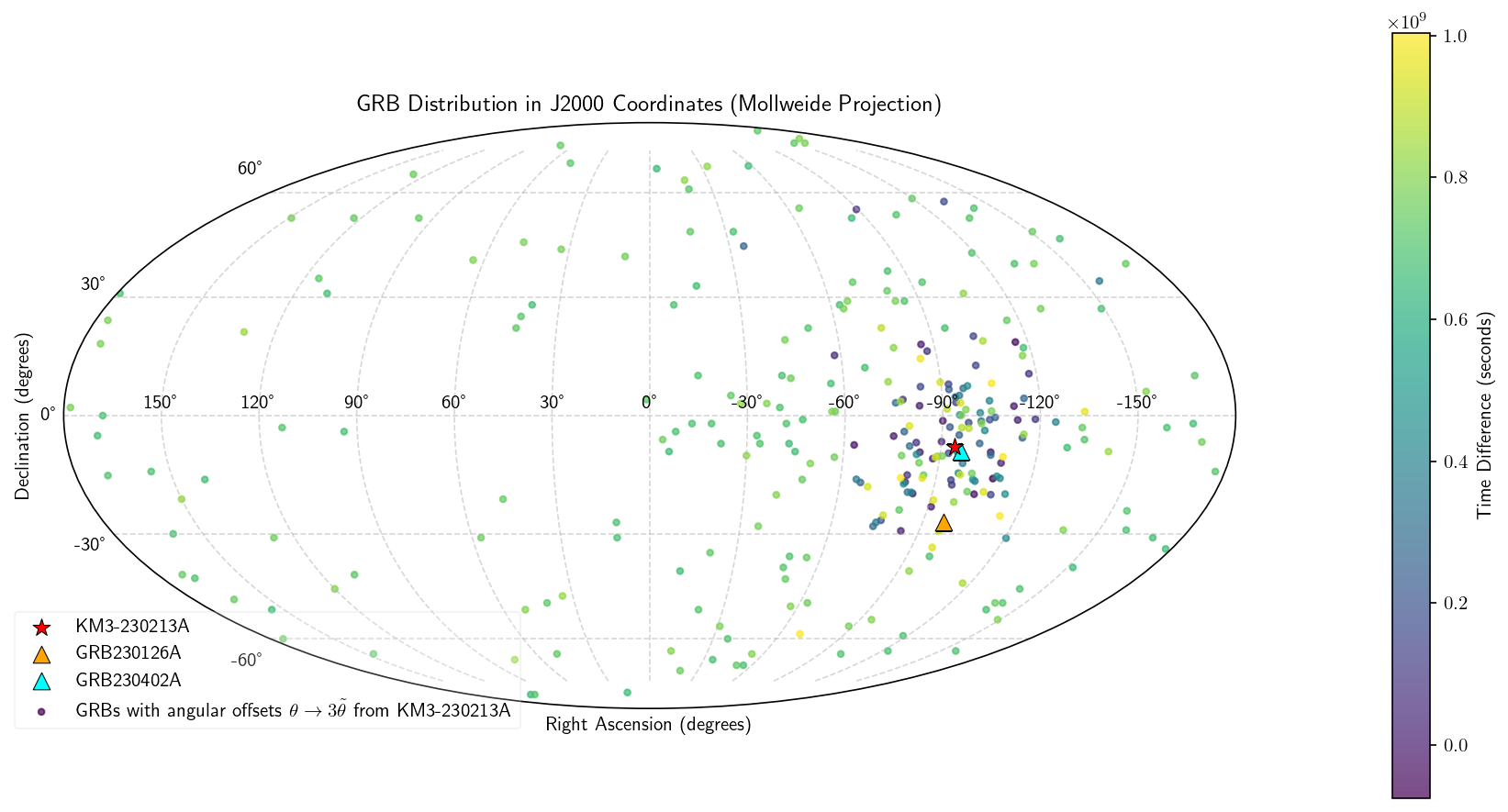}
    \caption{Mollweide projection showing the distribution of GRBs in J2000 coordinates, with colors representing time differences from KM3-230213A (marked by a red star). The orange and cyan triangles highlight GRB\,230126A and GRB\,230402A, respectively. All GRBs shown have angular offsets from $\theta$ to $3\tilde{\theta}$.
    }
    \label{fig:sky map2}
\end{figure}

Based on the absolute time difference from the neutrino event KM3-230213A, the positions of GRBs in the sky map are represented by points of different colors in Figures~\ref{fig:sky map1} and \ref{fig:sky map2}. It should be noted that some GRBs observed by the BeppoSAX GRBM detector have large positional errors~\cite{Frontera:2008sc}, resulting in varying degrees of GRB distribution across the sky map projection within the $\theta$ to $3\tilde{\theta}$ region. Both figures show that GRB events temporally closer to KM3-230213A are more concentrated near the neutrino event.

According to Table~\ref{Table 1} and Figure~\ref{fig:sky map1}, GRB\,920711A* exhibits the smallest angular separation from the neutrino event KM3-230213A, with an angular difference of $0.62\degree$. After incorporating the redshift and energy uncertainties of both the neutrino and GRBs into the analysis, the associated Lorentz-violating energy scale $E_{\rm LV}$ is determined to range from $(0.14\text{--}4.27)\times10^{17}~\rm GeV$, with a central value of $E_{\rm{LV}}= 5.7\times 10^{16}~\rm GeV$. Notably, this central value is consistent with the result reported in previous studies \cite{Huang:2022xto}.



In summary, we investigated the potential association between the neutrino event KM3-230213A and GRBs without pre-fixing the Lorentz-violation energy scale. By analyzing GRB candidates with angular differences from the neutrino within $\theta$ and in the range from $\theta$ to $3\tilde{\theta}$ respectively, we systematically considered spatiotemporal matching, energy consistency, and redshift constraints. Specifically, we incorporated angular, energy, and redshift uncertainties for both the neutrino and GRBs throughout the analysis. The key results are as follows:
\begin{itemize}
\item 
Multiple GRBs, including GRB\,090401B, show consistency with KM3-230213A at subluminal LV scales ($E_{\rm LV}\sim (3 \text{--} 10)\times10^{17}~\rm GeV$) under angular separation conditions of both within $\theta$ and from $\theta$ to $3\tilde{\theta}$. 
\item
GRB\,230126A ($E_{\rm{LV}}= 15.6^{+88.0}_{-11.6}\times 10^{19}~\rm GeV$, subluminal) and GRB\,230402A ($E_{\rm{LV}}= 5.6^{+31.5}_{-4.2}\times 10^{19}~\rm GeV$, superluminal) exhibit extreme LV parameter values, highlighting distinctive LV characteristics. 
\item
GRB\,920711A* exhibits the smallest angular separation ($0.62\degree$) from KM3-230213A, with an associated Lorentz-violating energy scale $E_{\rm LV}=(0.14 \text{--} 4.27)\times10^{17}~\rm GeV$, whose upper bound is consistent with the previously estimated $E_{\rm LV}\sim (3 \text{--} 10)\times10^{17}~\rm GeV$. 
\end{itemize}
These findings emphasize the significance of adopting flexible LV models in multi-messenger astrophysics studies and provide empirical constraints for scenarios in high-energy particle physics, contributing to a deeper understanding of fundamental physics within astrophysical contexts. 

Our strongest constraint yields a median value of $E_{\rm LV} \sim 1.56 \times 10^{20}~{\rm GeV}$, which is comparable to the limit of $E_{\rm QG,1} > 10\,E_{\rm Pl}$ ($\approx 1.2 \times 10^{20}~{\rm GeV}$) obtained by the LHAASO Collaboration from TeV gamma-ray observations of GRB\,221009A using
the same TOF method \cite{LHAASO:2024lub}. While both approaches employ the time-of-flight technique, they differ in several key aspects. The LHAASO result is based on TeV photons, which are subject to absorption by the extragalactic background light (EBL), limiting the effective redshift range of photon-based studies. In contrast, our neutrino-based approach exploits the $\sim$220~PeV energy of KM3-230213A, probing LV effects at energies several orders of magnitude higher than those accessible with photons, and
neutrinos propagate freely over cosmological distances without EBL absorption. Furthermore, photon-based TOF analyses require careful modeling of intrinsic source emission profiles to separate source-intrinsic spectral lags from LV-induced delays \cite{LHAASO:2024lub}, a systematic that does not affect our approach in the same way. On the other hand, the neutrino-based method is currently limited by the large energy uncertainty of the neutrino event and the difficulty of establishing firm neutrino-GRB associations. These complementary characteristics make neutrino- and photon-based TOF studies synergistic probes of LV, and future observations with improved angular resolution from KM3NeT and IceCube-Gen2 will significantly enhance the sensitivity of the neutrino-based approach \cite{KM3NeT2024potential, IceCubeGen2_2022}.

Establishing the physical reality of neutrino-GRB associations remains a key challenge for this type of study. Multi-wavelength spectral modeling of the GRB prompt and afterglow emission can be used to assess whether the physical conditions of a given burst --- such as the bulk Lorentz factor, radiation field density, and baryon loading --- are favorable for efficient neutrino production \cite{MuraseNagataki2006, Bustamante2015}. Additionally, comparing the observed neutrino energy with the theoretically expected neutrino fluence from the candidate GRB provides a consistency check on the assumed association. Looking ahead, next-generation neutrino telescopes including KM3NeT and IceCube-Gen2 will achieve significantly improved angular resolution, reducing positional uncertainties and chance coincidence probabilities, thereby enabling more robust identification of genuine neutrino-GRB associations \cite{KM3NeT2024potential, IceCubeGen2_2022}.

\begin{longtable}{lccccccc}
\caption{Table of associated GRBs indicating angular offsets within $\theta$ from neutrino KM3-230213A.}\\
\hline
\label{Table 1}
GRB Name & $\sigma_{\rm{GRB}}(\degree)$ & $\Delta\phi(\degree)$ & redshift z & $\Delta t$ ($10^3$s) & $K$ ($10^{18}$~TeV) & $E_{\text{LV}}$ ($10^{17}$~GeV) & $E_{\text{LV, 68\%}}$ ($10^{17}$GeV) \\
\hline
\endfirsthead
\hline
GRB Name & $\sigma_{\rm{GRB}}(\degree)$ & $\Delta\phi(\degree)$ & redshift z & $\Delta t$ ($10^3$s) & $K$ ($10^{18}$~TeV) & $E_{\text{LV}}$ ($10^{17}$~GeV) & $E_{\text{LV, 68\%}}$ ($10^{17}$GeV)  \\
\hline
\endhead
GRB\,240913B & 8.171 & 8.121 & 0.50$^\dagger$ & -50006.811 & 36584.175 & 10.974 & 2.661--82.520 \\
GRB\,240625B & 7.227 & 5.370 & 2.15$^\dagger$ & -43105.306 & 74129.961 & 54.172 & 13.979--358.903 \\
GRB\,220618B* & 26.197 & 14.447 & 0.50$^\dagger$ & 20674.873 & 36584.175 & 26.542 & 6.436--199.594 \\
GRB\,180227A* & 6.628 & 4.888 & 0.50$^\dagger$ & 156543.151 & 36584.175 & 3.506 & 0.850--26.361 \\
GRB\,170916A* & 16.405 & 15.526 & 2.15$^\dagger$ & 170670.559 & 74129.961 & 13.682 & 3.531--90.646 \\
GRB\,160804D* & 9.066 & 1.040 & 0.50$^\dagger$ & 205898.595 & 36584.175 & 2.665 & 0.646--20.042 \\
GRB\,130213A* & 9.695 & 4.753 & 2.15$^\dagger$ & 315459.179 & 74129.961 & 7.402 & 1.910--49.042 \\
GRB\,111024C* & 7.315 & 6.779 & 0.50$^\dagger$ & 356759.209 & 36584.175 & 1.538 & 0.373--11.567 \\
GRB\,100204C* & 15.619 & 13.458 & 0.50$^\dagger$ & 410935.248 & 36584.175 & 1.335 & 0.324--10.042 \\
GRB\,020212B & 72.000 & 67.433 & 2.15$^\dagger$ & 662694.292 & 74129.961 & 3.524 & 0.909--23.345 \\
GRB\,020119B & 33.000 & 32.822 & 2.15$^\dagger$ & 664836.583 & 74129.961 & 3.512 & 0.906--23.270 \\
GRB\,020102B & 33.000 & 15.015 & 2.15$^\dagger$ & 666245.331 & 74129.961 & 3.505 & 0.904--23.221 \\
GRB\,010913A & 57.000 & 44.404 & 0.50$^\dagger$ & 675836.894 & 36584.175 & 0.812 & 0.197--6.106 \\
GRB\,010605A & 49.000 & 41.961 & 0.50$^\dagger$ & 684498.981 & 36584.175 & 0.802 & 0.194--6.029 \\
GRB\,010602A & 53.000 & 14.907 & 2.15$^\dagger$ & 684767.985 & 74129.961 & 3.410 & 0.880--22.592 \\
GRB\,010407C & 54.000 & 53.619 & 2.15$^\dagger$ & 689568.046 & 74129.961 & 3.386 & 0.874--22.435 \\
GRB\,010222B & 53.000 & 35.661 & 0.50$^\dagger$ & 693375.790 & 36584.175 & 0.791 & 0.192--5.951 \\
GRB\,010212C & 55.000 & 38.338 & 2.15$^\dagger$ & 694228.780 & 74129.961 & 3.364 & 0.868--22.285 \\
GRB\,010123A & 33.000 & 32.129 & 2.15$^\dagger$ & 696034.692 & 74129.961 & 3.355 & 0.866--22.227 \\
GRB\,001218A & 56.000 & 26.897 & 2.15$^\dagger$ & 699145.845 & 74129.961 & 3.340 & 0.862--22.128 \\
GRB\,001216A & 53.000 & 49.861 & 0.50$^\dagger$ & 699275.011 & 36584.175 & 0.785 & 0.190--5.901 \\
GRB\,000718B & 18.000 & 17.300 & 2.15$^\dagger$ & 712297.784 & 74129.961 & 3.278 & 0.846--21.719 \\
GRB\,000704A & 31.000 & 30.196 & 2.15$^\dagger$ & 713571.440 & 74129.961 & 3.272 & 0.844--21.681 \\
GRB\,000626A & 40.000 & 18.654 & 0.50$^\dagger$ & 714198.564 & 36584.175 & 0.768 & 0.186--5.778 \\
GRB\,000618A & 39.000 & 29.267 & 2.15$^\dagger$ & 714932.473 & 74129.961 & 3.266 & 0.843--21.639 \\
GRB\,000502A & 47.000 & 31.127 & 2.15$^\dagger$ & 719013.209 & 74129.961 & 3.248 & 0.838--21.516 \\
GRB\,000208B & 25.000 & 14.553 & 2.15$^\dagger$ & 726249.885 & 74129.961 & 3.215 & 0.830--21.302 \\
GRB\,000107B & 48.000 & 26.844 & 2.15$^\dagger$ & 728973.654 & 74129.961 & 3.203 & 0.827--21.222 \\
GRB\,990717A & 30.000 & 10.862 & 2.15$^\dagger$ & 744019.478 & 74129.961 & 3.138 & 0.810--20.793 \\
GRB\,990630A & 31.000 & 26.813 & 2.15$^\dagger$ & 745493.855 & 74129.961 & 3.132 & 0.808--20.752 \\
GRB\,990624A & 32.000 & 13.553 & 2.15$^\dagger$ & 745992.197 & 74129.961 & 3.130 & 0.808--20.738 \\
GRB\,990601A & 49.000 & 26.167 & 2.15$^\dagger$ & 748001.720 & 74129.961 & 3.122 & 0.806--20.683 \\
GRB\,980904A & 48.000 & 42.451 & 1.88$^\ddagger$ & 771359.649 & 71725.754 & 2.678 & 0.683--18.005 \\
GRB\,980830A & 46.000 & 41.246 & 0.50$^\dagger$ & 771762.893 & 36584.175 & 0.711 & 0.172--5.347 \\
GRB\,980820A & 26.000 & 15.015 & 2.15$^\dagger$ & 772595.497 & 74129.961 & 3.022 & 0.780--20.024 \\
GRB\,980606A & 35.000 & 26.085 & 0.50$^\dagger$ & 779127.279 & 36584.175 & 0.704 & 0.171--5.296 \\
GRB\,980516A & 15.000 & 6.498 & 2.15$^\dagger$ & 780933.197 & 74129.961 & 2.990 & 0.772--19.810 \\
GRB\,980118A & 39.000 & 7.982 & 2.15$^\dagger$ & 791126.586 & 74129.961 & 2.952 & 0.762--19.555 \\
GRB\,971223B & 39.000 & 25.851 & 2.15$^\dagger$ & 793381.314 & 74129.961 & 2.943 & 0.759--19.500 \\
GRB\,971130A & 40.000 & 33.144 & 2.15$^\dagger$ & 795329.233 & 74129.961 & 2.936 & 0.758--19.452 \\
GRB\,970812A & 28.000 & 26.606 & 2.15$^\dagger$ & 804893.853 & 74129.961 & 2.901 & 0.749--19.221 \\
GRB\,970517C & 47.000 & 36.102 & 0.50$^\dagger$ & 812377.448 & 36584.175 & 0.676 & 0.164--5.080 \\
GRB\,970504A & 44.000 & 15.215 & 0.50$^\dagger$ & 813517.410 & 36584.175 & 0.675 & 0.164--5.073 \\
GRB\,970303A & 42.000 & 35.848 & 0.50$^\dagger$ & 818901.597 & 36584.175 & 0.670 & 0.162--5.039 \\
GRB\,970104A & 55.000 & 39.734 & 1.88$^\ddagger$ & 823902.639 & 71725.754 & 2.507 & 0.639--16.857 \\
GRB\,961208B & 50.200 & 46.762 & 2.15$^\dagger$ & 826179.582 & 74129.961 & 2.826 & 0.729--18.725 \\
GRB\,960929A & 42.000 & 38.705 & 2.15$^\dagger$ & 832272.033 & 74129.961 & 2.806 & 0.724--18.588 \\
GRB\,960810B & 33.000 & 8.242 & 1.88$^\ddagger$ & 836590.839 & 71725.754 & 2.469 & 0.629--16.601 \\
GRB\,940214C* & 3.170 & 2.777 & 1.88$^\ddagger$ & 915025.340 & 71725.754 & 2.258 & 0.575--15.178 \\
GRB\,940124A* & 14.990 & 9.561 & 0.50$^\dagger$ & 916831.138 & 36584.175 & 0.599 & 0.145--4.501 \\
GRB\,940101A* & 10.120 & 1.312 & 2.15$^\dagger$ & 918838.510 & 74129.961 & 2.541 & 0.656--16.837 \\
GRB\,930228A* & 18.570 & 14.805 & 0.50$^\dagger$ & 945363.900 & 36584.175 & 0.580 & 0.141--4.365 \\
GRB\,930201A* & 9.350 & 6.098 & 1.88$^\ddagger$ & 947694.577 & 71725.754 & 2.180 & 0.556--14.655 \\
GRB\,920711A* & 13.070 & 0.622 & 0.50$^\dagger$ & 965400.525 & 36584.175 & 0.568 & 0.138--4.274 \\
\hline
\addlinespace[6pt]
\multicolumn{8}{c}{\parbox{0.8\textwidth}{\footnotesize * denotes GRBs without GCN-style names (auto-generated by GRBweb).  $^\ddagger$ shows estimated average redshift values $z=1.88$ for GRBs lacking redshift/duration data~\cite{10.3389/fspas.2023.1124317}, and $^\dagger$ denotes average redshift estimates ($z=2.15$ for long bursts; $z=0.5$ for short bursts).}}
\end{longtable}

\begin{longtable}{lccccccc}
\caption{Table of Associated GRBs indicating angular offsets between $\theta$ and $3\tilde{\theta}$ from neutrino KM3-230213A.} \\
\toprule
\label{Table 2}
GRB Name & $\sigma_{\rm{GRB}}(\degree)$ & $\Delta\phi(\degree)$ & redshift z & $\Delta t$ ($10^3$s) & $K$ ($10^{18}$~TeV) & $E_{\text{LV}}$ ($10^{17}$~GeV) & $E_{\text{LV, 68\%}}$ ($10^{17}$GeV) \\
\midrule
\endfirsthead
\toprule
GRB Name & $\sigma_{\rm{GRB}}(\degree)$ & $\Delta\phi(\degree)$ & redshift z & $\Delta t$ ($10^3$s) & $K$ ($10^{18}$~TeV) & $E_{\text{LV}}$ ($10^{17}$~GeV) & $E_{\text{LV, 68\%}}$ ($10^{17}$GeV)\\
\midrule
\endhead
GRB\,250702C & 8.773 & 15.530 & 0.50$^\dagger$ & -75216.763 & 36584.175 & 7.296 & 1.769--54.863 \\
GRB\,240619B & 12.745 & 33.877 & 2.15$^\dagger$ & -42544.439 & 74129.961 & 54.886 & 14.163--363.634 \\
GRB\,240215C* & 8.134 & 14.847 & 2.15$^\dagger$ & -31762.115 & 74129.961 & 73.518 & 18.971--487.077 \\
GRB\,231123A & 6.788 & 15.574 & 2.15$^\dagger$ & -24456.436 & 74129.961 & 95.480 & 24.638--632.578 \\
GRB\,230930A & 4.767 & 11.228 & 2.15$^\dagger$ & -19860.217 & 74129.961 & 117.576 & 30.341--778.975 \\
GRB\,230806A* & 13.691 & 30.905 & 0.50$^\dagger$ & -15043.545 & 36584.175 & 36.478 & 8.845--274.309 \\
GRB\,230709A* & 2.855 & 7.749 & 2.15$^\dagger$ & -12670.767 & 74129.961 & 184.290 & 47.556--1220.969 \\
GRB\,230402A & 0.400 & 2.351 & 2.15$^\dagger$ & -4169.747 & 74129.961 & 560.009 & 144.510--3710.203 \\
GRB\,230129A & 13.086 & 20.314 & 0.50$^\dagger$ & 1278.686 & 36584.175 & 429.161 & 104.058--3227.207 \\
GRB\,230126A & 7.496 & 19.343 & 2.15$^\dagger$ & 1493.774 & 74129.961 & 1563.218 & 403.388--10356.727 \\
GRB\,230112B* & 9.025 & 10.619 & 2.15$^\dagger$ & 2697.781 & 74129.961 & 865.561 & 223.358--5734.568 \\
GRB\,230112A & 9.485 & 19.243 & 0.50$^\dagger$ & 2739.855 & 36584.175 & 200.289 & 48.564--1506.132 \\
GRB\,221014A & 12.175 & 26.936 & 0.50$^\dagger$ & 10543.622 & 36584.175 & 52.047 & 12.620--391.382 \\
GRB\,220927A & 5.100 & 15.485 & 2.15$^\dagger$ & 11994.062 & 74129.961 & 194.687 & 50.239--1289.856 \\
GRB\,220827A & 11.166 & 19.847 & 2.15$^\dagger$ & 14615.207 & 74129.961 & 159.772 & 41.229--1058.528 \\
GRB\,220825A & 6.157 & 6.983 & 2.15$^\dagger$ & 14858.346 & 74129.961 & 157.157 & 40.554--1041.207 \\
GRB\,220823A & 10.342 & 15.273 & 0.50$^\dagger$ & 15025.676 & 36584.175 & 36.522 & 8.855--274.636 \\
GRB\,220513A & 8.220 & 23.536 & 0.50$^\dagger$ & 23776.240 & 36584.175 & 23.080 & 5.596--173.559 \\
GRB\,220510A* & 6.673 & 15.370 & 0.50$^\dagger$ & 24097.542 & 36584.175 & 22.773 & 5.522--171.245 \\
GRB\,210923A & 2.900$^a$ & 4.460 & 1.88$^\ddagger$ & 43853.232 & 71725.754 & 47.105 & 12.005--316.696 \\
GRB\,210326A* & 14.376 & 42.618 & 0.50$^\dagger$ & 59529.263 & 36584.175 & 9.218 & 2.235--69.320 \\
GRB\,201111A & 8.449 & 16.061 & 2.15$^\dagger$ & 71170.262 & 74129.961 & 32.810 & 8.467--217.375 \\
GRB\,200824A & 10.264 & 29.588 & 2.15$^\dagger$ & 77972.471 & 74129.961 & 29.948 & 7.728--198.411 \\
GRB\,200710A & 10.332 & 20.177 & 0.50$^\dagger$ & 81908.681 & 36584.175 & 6.700 & 1.624--50.380 \\
GRB\,200707A* & 4.925 & 15.028 & 2.15$^\dagger$ & 82164.765 & 74129.961 & 28.420 & 7.334--188.288 \\
GRB\,200706A & 23.984 & 62.548 & 0.50$^\dagger$ & 82237.479 & 36584.175 & 6.673 & 1.618--50.179 \\
GRB\,200626A & 9.750 & 24.899 & 2.15$^\dagger$ & 83112.317 & 74129.961 & 28.096 & 7.250--186.141 \\
GRB\,200506B & 10.880 & 12.019 & 2.15$^\dagger$ & 87509.829 & 74129.961 & 26.684 & 6.886--176.787 \\
GRB\,200103B* & 8.591 & 25.063 & 2.15$^\dagger$ & 98181.864 & 74129.961 & 23.783 & 6.137--157.571 \\
GRB\,191225A & 5.355 & 9.620 & 2.15$^\dagger$ & 98992.297 & 74129.961 & 23.589 & 6.087--156.281 \\
GRB\,190502A* & 3.746 & 11.091 & 2.15$^\dagger$ & 119481.318 & 74129.961 & 19.544 & 5.043--129.481 \\
GRB\,190204B* & 6.438 & 8.368 & 2.15$^\dagger$ & 126958.459 & 74129.961 & 18.393 & 4.746--121.856 \\
GRB\,180722B* & 6.996 & 18.500 & 2.15$^\dagger$ & 144023.014 & 74129.961 & 16.213 & 4.184--107.418 \\
GRB\,180525A* & 10.951 & 29.108 & 0.50$^\dagger$ & 149031.528 & 36584.175 & 3.682 & 0.893--27.689 \\
GRB\,180404D* & 10.042 & 21.634 & 0.50$^\dagger$ & 153377.777 & 36584.175 & 3.578 & 0.868--26.905 \\
GRB\,180404C & 5.667 & 15.743 & 1.88$^\ddagger$ & 153372.880 & 71725.754 & 13.468 & 3.433--90.551 \\
GRB\,171025B* & 6.793 & 7.067 & 2.15$^\dagger$ & 167325.473 & 74129.961 & 13.955 & 3.601--92.458 \\
GRB\,170921B & 37.500 & 71.922 & 2.15$^\dagger$ & 170284.479 & 74129.961 & 13.713 & 3.539--90.852 \\
GRB\,170912C* & 16.349 & 18.425 & 0.50$^\dagger$ & 170991.516 & 36584.175 & 3.209 & 0.778--24.133 \\
GRB\,170621A* & 8.669 & 25.449 & 2.15$^\dagger$ & 178180.045 & 74129.961 & 13.105 & 3.382--86.826 \\
GRB\,170610A & 0.233 & 2.356 & 1.88$^\ddagger$ & 179123.630 & 71725.754 & 11.532 & 2.939--77.534 \\
GRB\,160826B* & 11.115 & 12.184 & 0.50$^\dagger$ & 204000.359 & 36584.175 & 2.690 & 0.652--20.228 \\
GRB\,160226A* & 3.419 & 5.317 & 2.15$^\dagger$ & 219727.350 & 74129.961 & 10.627 & 2.742--70.408 \\
GRB\,150923D* & 5.833 & 14.471 & 2.15$^\dagger$ & 233198.636 & 74129.961 & 10.013 & 2.584--66.341 \\
GRB\,150613A* & 5.183 & 13.889 & 2.15$^\dagger$ & 242061.148 & 74129.961 & 9.647 & 2.489--63.912 \\
GRB\,150303A* & 10.582 & 20.493 & 2.15$^\dagger$ & 250865.639 & 74129.961 & 9.308 & 2.402--61.669 \\
GRB\,150213A & 2.492 & 3.110 & 2.15$^\dagger$ & 252465.301 & 74129.961 & 9.249 & 2.387--61.278 \\
GRB\,141112B* & 6.259 & 16.349 & 2.15$^\dagger$ & 260429.092 & 74129.961 & 8.966 & 2.314--59.404 \\
GRB\,140627A* & 12.370 & 28.571 & 2.15$^\dagger$ & 272389.133 & 74129.961 & 8.573 & 2.212--56.796 \\
GRB\,140516B* & 11.449 & 24.068 & 2.15$^\dagger$ & 275992.163 & 74129.961 & 8.461 & 2.183--56.055 \\
GRB\,140416A & 25.800 & 74.059 & 2.15$^\dagger$ & 278639.414 & 74129.961 & 8.380 & 2.163--55.522 \\
GRB\,140404E* & 3.637 & 7.496 & 2.15$^\dagger$ & 279603.633 & 74129.961 & 8.351 & 2.155--55.331 \\
GRB\,140113A* & 8.114 & 21.621 & 2.15$^\dagger$ & 286663.980 & 74129.961 & 8.146 & 2.102--53.968 \\
GRB\,131102A* & 16.380 & 27.794 & 2.15$^\dagger$ & 292846.865 & 74129.961 & 7.974 & 2.058--52.828 \\
GRB\,130802A* & 8.297 & 13.894 & 0.50$^\dagger$ & 300786.297 & 36584.175 & 1.824 & 0.442--13.719 \\
GRB\,130510A* & 3.963 & 11.462 & 2.15$^\dagger$ & 308031.212 & 74129.961 & 7.581 & 1.956--50.224 \\
GRB\,130217A* & 6.454 & 14.798 & 2.15$^\dagger$ & 315132.341 & 74129.961 & 7.410 & 1.912--49.092 \\
GRB\,120916B* & 11.707 & 16.491 & 0.50$^\dagger$ & 328490.074 & 36584.175 & 1.671 & 0.405--12.562 \\
GRB\,120218B* & 3.954 & 9.896 & 2.15$^\dagger$ & 346704.201 & 74129.961 & 6.735 & 1.738--44.622 \\
GRB\,120111A* & 5.092 & 12.842 & 2.15$^\dagger$ & 350006.605 & 74129.961 & 6.672 & 1.722--44.201 \\
GRB\,111231A* & 30.180 & 71.481 & 2.15$^\dagger$ & 350907.659 & 74129.961 & 6.654 & 1.717--44.087 \\
GRB\,111124A* & 6.365 & 12.432 & 2.15$^\dagger$ & 354131.561 & 74129.961 & 6.594 & 1.702--43.686 \\
GRB\,110624A* & 13.374 & 29.750 & 2.15$^\dagger$ & 367299.147 & 74129.961 & 6.357 & 1.641--42.120 \\
GRB\,110526A* & 4.884 & 11.753 & 0.50$^\dagger$ & 369821.269 & 36584.175 & 1.484 & 0.360--11.158 \\
GRB\,110509B* & 10.326 & 26.738 & 0.50$^\dagger$ & 371310.755 & 36584.175 & 1.478 & 0.358--11.114 \\
GRB\,110407B* & 2.532 & 5.158 & 2.15$^\dagger$ & 374030.394 & 74129.961 & 6.243 & 1.611--41.362 \\
GRB\,100810A* & 10.636 & 30.975 & 2.15$^\dagger$ & 394848.378 & 74129.961 & 5.914 & 1.526--39.181 \\
GRB\,100301A* & 6.531 & 17.382 & 0.50$^\dagger$ & 408843.573 & 36584.175 & 1.342 & 0.325--10.093 \\
GRB\,100223A & 5.611 & 15.356 & 0.50$^\dagger$ & 409358.321 & 36584.175 & 1.341 & 0.325--10.081 \\
GRB\,091230B* & 10.493 & 11.130 & 2.15$^\dagger$ & 414097.365 & 74129.961 & 5.639 & 1.455--37.360 \\
GRB\,091126A & 8.700 & 15.700 & 0.50$^\dagger$ & 417028.646 & 36584.175 & 1.316 & 0.319--9.895 \\
GRB\,091122A* & 15.983 & 33.794 & 2.15$^\dagger$ & 417388.953 & 74129.961 & 5.595 & 1.444--37.065 \\
GRB\,090809B & 2.506 & 8.016 & 2.15$^\dagger$ & 426390.516 & 74129.961 & 5.476 & 1.413--36.283 \\
GRB\,090616A & 8.232 & 9.635 & 0.50$^\dagger$ & 431127.068 & 36584.175 & 1.273 & 0.309--9.572 \\
GRB\,090426C & 3.790 & 11.620 & 2.15$^\dagger$ & 435487.399 & 74129.961 & 5.362 & 1.384--35.525 \\
GRB\,090401B & 0.00007 & 1.408 & 2.15$^\dagger$ & 437676.088 & 74129.961 & 5.335 & 1.377--35.347 \\
GRB\,081017B* & 8.183 & 16.183 & 2.15$^\dagger$ & 452008.468 & 74129.961 & 5.166 & 1.333--34.226 \\
GRB\,080818B* & 5.460 & 15.665 & 2.15$^\dagger$ & 457151.763 & 74129.961 & 5.108 & 1.318--33.841 \\
GRB\,080817B* & 6.328 & 16.599 & 2.15$^\dagger$ & 457257.585 & 74129.961 & 5.107 & 1.318--33.833 \\
GRB\,080808C* & 2.900$^a$ & 7.007 & 2.15$^\dagger$ & 458030.880 & 74129.961 & 5.098 & 1.316--33.776 \\
GRB\,080807A* & 4.878 & 10.935 & 2.15$^\dagger$ & 458097.979 & 74129.961 & 5.097 & 1.315--33.771 \\
GRB\,080723C* & 7.437 & 21.922 & 0.50$^\dagger$ & 459400.889 & 36584.175 & 1.195 & 0.290--8.983 \\
GRB\,020416A & 33.000 & 97.341 & 2.15$^\dagger$ & 657307.419 & 74129.961 & 3.553 & 0.917--23.536 \\
GRB\,020414A & 49.000 & 89.656 & 0.50$^\dagger$ & 657502.169 & 36584.175 & 0.835 & 0.202--6.276 \\
GRB\,020401A & 36.000 & 86.553 & 1.88$^\ddagger$ & 658570.880 & 71725.754 & 3.137 & 0.799--21.088 \\
GRB\,020327A & 44.000 & 60.209 & 2.15$^\dagger$ & 659057.860 & 74129.961 & 3.543 & 0.914--23.474 \\
GRB\,020320A & 48.000 & 50.848 & 1.88$^\ddagger$ & 659647.323 & 71725.754 & 3.132 & 0.798--21.054 \\
GRB\,020308A & 37.000 & 87.119 & 2.15$^\dagger$ & 660683.394 & 74129.961 & 3.534 & 0.912--23.416 \\
GRB\,020209A & 42.000 & 96.599 & 2.15$^\dagger$ & 663010.016 & 74129.961 & 3.522 & 0.909--23.334 \\
GRB\,020119C & 46.000 & 80.825 & 2.15$^\dagger$ & 664794.364 & 74129.961 & 3.513 & 0.906--23.271 \\
GRB\,020113A & 41.000 & 60.450 & 2.15$^\dagger$ & 665363.562 & 74129.961 & 3.510 & 0.906--23.251 \\
GRB\,020102A & 50.000 & 79.876 & 2.15$^\dagger$ & 666245.944 & 74129.961 & 3.505 & 0.904--23.221 \\
GRB\,011231A & 53.000 & 80.443 & 2.15$^\dagger$ & 666481.325 & 74129.961 & 3.504 & 0.904--23.212 \\
GRB\,011230A & 39.000 & 84.573 & 2.15$^\dagger$ & 666552.635 & 74129.961 & 3.503 & 0.904--23.210 \\
GRB\,011222A & 32.000 & 34.361 & 2.15$^\dagger$ & 667233.709 & 74129.961 & 3.500 & 0.903--23.186 \\
GRB\,011217A & 53.000 & 121.221 & 1.88$^\ddagger$ & 667661.625 & 71725.754 & 3.094 & 0.789--20.801 \\
GRB\,011212A & 51.000 & 73.760 & 2.15$^\dagger$ & 668130.058 & 74129.961 & 3.495 & 0.902--23.155 \\
GRB\,011126A & 31.000 & 48.750 & 2.15$^\dagger$ & 669481.970 & 74129.961 & 3.488 & 0.900--23.108 \\
GRB\,011104A & 32.000 & 94.945 & 2.15$^\dagger$ & 671382.890 & 74129.961 & 3.478 & 0.898--23.043 \\
GRB\,010908A & 53.000 & 148.198 & 2.15$^\dagger$ & 676279.008 & 74129.961 & 3.453 & 0.891--22.876 \\
GRB\,010818B & 41.000 & 72.592 & 2.15$^\dagger$ & 678082.753 & 74129.961 & 3.444 & 0.889--22.815 \\
GRB\,010723A & 83.000 & 165.584 & 2.15$^\dagger$ & 680344.018 & 74129.961 & 3.432 & 0.886--22.739 \\
GRB\,010721A & 29.000 & 34.119 & 2.15$^\dagger$ & 680563.202 & 74129.961 & 3.431 & 0.885--22.732 \\
GRB\,010710B & 25.000 & 63.353 & 2.15$^\dagger$ & 681442.960 & 74129.961 & 3.427 & 0.884--22.703 \\
GRB\,010705A & 20.000 & 33.598 & 2.15$^\dagger$ & 681944.561 & 74129.961 & 3.424 & 0.884--22.686 \\
GRB\,010623A & 44.000 & 131.579 & 2.15$^\dagger$ & 682983.938 & 74129.961 & 3.419 & 0.882--22.651 \\
GRB\,010619A & 20.000 & 59.805 & 2.15$^\dagger$ & 683331.507 & 74129.961 & 3.417 & 0.882--22.640 \\
GRB\,010618B & 29.000 & 29.808 & 2.15$^\dagger$ & 683370.616 & 74129.961 & 3.417 & 0.882--22.639 \\
GRB\,010616A & 30.000 & 46.863 & 2.15$^\dagger$ & 683595.528 & 74129.961 & 3.416 & 0.881--22.631 \\
GRB\,010611A & 41.000 & 98.870 & 2.15$^\dagger$ & 684005.912 & 74129.961 & 3.414 & 0.881--22.618 \\
GRB\,010528A & 36.000 & 41.318 & 2.15$^\dagger$ & 685230.988 & 74129.961 & 3.408 & 0.879--22.577 \\
GRB\,010518B & 40.000 & 56.204 & 2.15$^\dagger$ & 686076.740 & 74129.961 & 3.404 & 0.878--22.549 \\
GRB\,010515A & 27.000 & 71.608 & 2.15$^\dagger$ & 686357.781 & 74129.961 & 3.402 & 0.878--22.540 \\
GRB\,010514A & 38.000 & 66.331 & 2.15$^\dagger$ & 686392.474 & 74129.961 & 3.402 & 0.878--22.539 \\
GRB\,010510A & 46.000 & 47.612 & 2.15$^\dagger$ & 686795.647 & 74129.961 & 3.400 & 0.877--22.526 \\
GRB\,010505B & 42.000 & 97.628 & 2.15$^\dagger$ & 687187.863 & 74129.961 & 3.398 & 0.877--22.513 \\
GRB\,010505A & 39.000 & 55.976 & 2.15$^\dagger$ & 687207.025 & 74129.961 & 3.398 & 0.877--22.512 \\
GRB\,010430A & 39.000 & 85.800 & 1.88$^\ddagger$ & 687579.203 & 71725.754 & 3.004 & 0.766--20.199 \\
GRB\,010427C & 24.000 & 28.402 & 2.15$^\dagger$ & 687873.767 & 74129.961 & 3.395 & 0.876--22.490 \\
GRB\,010420A & 33.000 & 64.537 & 2.15$^\dagger$ & 688444.737 & 74129.961 & 3.392 & 0.875--22.472 \\
GRB\,010415B & 37.000 & 99.458 & 1.88$^\ddagger$ & 688889.110 & 71725.754 & 2.999 & 0.764--20.160 \\
GRB\,010411A & 52.000 & 81.112 & 2.15$^\dagger$ & 689256.412 & 74129.961 & 3.388 & 0.874--22.445 \\
GRB\,010407A & 34.000 & 45.828 & 2.15$^\dagger$ & 689623.196 & 74129.961 & 3.386 & 0.874--22.433 \\
GRB\,010406A & 33.000 & 78.035 & 1.88$^\ddagger$ & 689732.437 & 71725.754 & 2.995 & 0.763--20.136 \\
GRB\,010404A & 37.000 & 80.933 & 2.15$^\dagger$ & 689878.144 & 74129.961 & 3.385 & 0.873--22.425 \\
GRB\,010330A & 59.000 & 71.182 & 2.15$^\dagger$ & 690288.510 & 74129.961 & 3.383 & 0.873--22.412 \\
GRB\,010321A & 36.000 & 106.931 & 2.15$^\dagger$ & 691074.516 & 74129.961 & 3.379 & 0.872--22.386 \\
GRB\,010317A & 24.000 & 60.935 & 2.15$^\dagger$ & 691440.525 & 74129.961 & 3.377 & 0.871--22.374 \\
GRB\,010309A & 29.000 & 75.168 & 2.15$^\dagger$ & 692109.781 & 74129.961 & 3.374 & 0.871--22.353 \\
GRB\,010308A & 30.000 & 67.347 & 2.15$^\dagger$ & 692185.076 & 74129.961 & 3.374 & 0.871--22.350 \\
GRB\,010307A & 35.000 & 64.144 & 2.15$^\dagger$ & 692268.552 & 74129.961 & 3.373 & 0.870--22.348 \\
GRB\,010226B & 65.000 & 91.204 & 2.15$^\dagger$ & 693028.180 & 74129.961 & 3.369 & 0.869--22.323 \\
GRB\,010209A & 44.000 & 90.416 & 1.88$^\ddagger$ & 694555.554 & 71725.754 & 2.974 & 0.758--19.996 \\
GRB\,010208C & 48.000 & 57.865 & 2.15$^\dagger$ & 694608.635 & 74129.961 & 3.362 & 0.867--22.272 \\
GRB\,010208B & 35.000 & 77.438 & 2.15$^\dagger$ & 694645.710 & 74129.961 & 3.362 & 0.867--22.271 \\
GRB\,010203A & 33.000 & 54.298 & 1.88$^\ddagger$ & 695075.967 & 71725.754 & 2.972 & 0.757--19.981 \\
GRB\,010126C & 40.000 & 99.865 & 2.15$^\dagger$ & 695710.203 & 74129.961 & 3.356 & 0.866--22.237 \\
GRB\,010121A & 40.000 & 54.800 & 2.15$^\dagger$ & 696152.897 & 74129.961 & 3.354 & 0.866--22.223 \\
GRB\,010114A & 30.000 & 71.675 & 2.15$^\dagger$ & 696807.181 & 74129.961 & 3.351 & 0.865--22.202 \\
GRB\,010111A & 38.000 & 73.826 & 2.15$^\dagger$ & 697012.404 & 74129.961 & 3.350 & 0.865--22.196 \\
GRB\,001215A & 29.000 & 55.037 & 2.15$^\dagger$ & 699394.818 & 74129.961 & 3.339 & 0.862--22.120 \\
GRB\,001214A & 64.000 & 90.449 & 2.15$^\dagger$ & 699477.465 & 74129.961 & 3.338 & 0.861--22.117 \\
GRB\,001206A & 33.000 & 49.942 & 2.15$^\dagger$ & 700155.591 & 74129.961 & 3.335 & 0.861--22.096 \\
GRB\,001201A & 44.000 & 57.395 & 2.15$^\dagger$ & 700574.877 & 74129.961 & 3.333 & 0.860--22.083 \\
GRB\,001118A & 22.000 & 52.393 & 2.15$^\dagger$ & 701715.428 & 74129.961 & 3.328 & 0.859--22.047 \\
GRB\,001115A & 21.000 & 59.751 & 2.15$^\dagger$ & 701953.812 & 74129.961 & 3.327 & 0.858--22.039 \\
GRB\,001106A & 40.000 & 93.435 & 0.50$^\dagger$ & 702719.945 & 36584.175 & 0.781 & 0.189--5.872 \\
GRB\,001101A & 27.000 & 54.683 & 2.15$^\dagger$ & 703135.462 & 74129.961 & 3.321 & 0.857--22.002 \\
GRB\,000924A & 40.000 & 51.877 & 2.15$^\dagger$ & 706437.376 & 74129.961 & 3.305 & 0.853--21.899 \\
GRB\,000922A & 33.000 & 41.801 & 2.15$^\dagger$ & 706593.322 & 74129.961 & 3.305 & 0.853--21.895 \\
GRB\,000916A & 20.000 & 37.876 & 2.15$^\dagger$ & 707137.311 & 74129.961 & 3.302 & 0.852--21.878 \\
GRB\,000915A & 40.000 & 89.987 & 2.15$^\dagger$ & 707260.714 & 74129.961 & 3.302 & 0.852--21.874 \\
GRB\,000903C & 17.000 & 39.440 & 2.15$^\dagger$ & 708255.870 & 74129.961 & 3.297 & 0.851--21.843 \\
GRB\,000830B & 22.000 & 56.756 & 2.15$^\dagger$ & 708615.465 & 74129.961 & 3.295 & 0.850--21.832 \\
GRB\,000828B & 47.000 & 60.123 & 2.15$^\dagger$ & 708747.279 & 74129.961 & 3.295 & 0.850--21.828 \\
GRB\,000828A & 41.000 & 89.817 & 2.15$^\dagger$ & 708765.674 & 74129.961 & 3.295 & 0.850--21.828 \\
GRB\,000819A & 38.000 & 81.084 & 2.15$^\dagger$ & 709578.077 & 74129.961 & 3.291 & 0.849--21.803 \\
GRB\,000723A & 41.000 & 74.086 & 2.15$^\dagger$ & 711902.207 & 74129.961 & 3.280 & 0.846--21.731 \\
GRB\,000629A & 50.000 & 61.358 & 2.15$^\dagger$ & 714012.515 & 74129.961 & 3.270 & 0.844--21.667 \\
GRB\,000627A & 29.000 & 51.852 & 2.15$^\dagger$ & 714170.084 & 74129.961 & 3.270 & 0.844--21.662 \\
GRB\,000621A & 68.000 & 150.702 & 2.15$^\dagger$ & 714698.162 & 74129.961 & 3.267 & 0.843--21.646 \\
GRB\,000613A & 34.000 & 68.340 & 2.15$^\dagger$ & 715394.708 & 74129.961 & 3.264 & 0.842--21.625 \\
GRB\,000525A & 33.000 & 76.474 & 2.15$^\dagger$ & 717016.099 & 74129.961 & 3.257 & 0.840--21.576 \\
GRB\,000502B & 47.000 & 127.068 & 2.15$^\dagger$ & 718977.498 & 74129.961 & 3.248 & 0.838--21.518 \\
GRB\,000402A & 55.000 & 134.953 & 2.15$^\dagger$ & 721606.862 & 74129.961 & 3.236 & 0.835--21.439 \\
GRB\,000214B & 50.000 & 111.075 & 0.50$^\dagger$ & 725743.954 & 36584.175 & 0.756 & 0.183--5.686 \\
GRB\,000114A & 34.000 & 94.845 & 2.15$^\dagger$ & 728410.821 & 74129.961 & 3.206 & 0.827--21.239 \\
GRB\,000110A & 42.000 & 113.306 & 2.15$^\dagger$ & 728771.887 & 74129.961 & 3.204 & 0.827--21.228 \\
GRB\,991212A & 52.000 & 136.547 & 0.50$^\dagger$ & 731269.409 & 36584.175 & 0.750 & 0.182--5.643 \\
GRB\,991209B & 43.000 & 95.783 & 1.88$^\ddagger$ & 731484.319 & 71725.754 & 2.824 & 0.720--18.986 \\
GRB\,991205B & 44.000 & 71.877 & 2.15$^\dagger$ & 731818.323 & 74129.961 & 3.191 & 0.823--21.140 \\
GRB\,991201A & 51.000 & 141.775 & 2.15$^\dagger$ & 732162.338 & 74129.961 & 3.189 & 0.823--21.130 \\
GRB\,991128A & 46.000 & 87.061 & 2.15$^\dagger$ & 732481.973 & 74129.961 & 3.188 & 0.823--21.121 \\
GRB\,991124A & 38.000 & 93.743 & 2.15$^\dagger$ & 732825.291 & 74129.961 & 3.186 & 0.822--21.111 \\
GRB\,991119A & 48.000 & 102.281 & 2.15$^\dagger$ & 733228.977 & 74129.961 & 3.185 & 0.822--21.099 \\
GRB\,991105B & 46.000 & 88.076 & 0.50$^\dagger$ & 734452.641 & 36584.175 & 0.747 & 0.181--5.619 \\
GRB\,991101A & 33.000 & 70.303 & 2.15$^\dagger$ & 734823.031 & 74129.961 & 3.178 & 0.820--21.054 \\
GRB\,991028A* & 13.260 & 20.459 & 2.15$^\dagger$ & 735150.355 & 74129.961 & 3.176 & 0.820--21.044 \\
GRB\,991026B & 41.000 & 67.350 & 2.15$^\dagger$ & 735308.076 & 74129.961 & 3.176 & 0.819--21.040 \\
GRB\,990917B & 4.700 & 10.869 & 2.15$^\dagger$ & 738672.113 & 74129.961 & 3.161 & 0.816--20.944 \\
GRB\,990903A & 43.000 & 126.310 & 2.15$^\dagger$ & 739913.682 & 74129.961 & 3.156 & 0.814--20.909 \\
GRB\,990827A & 47.000 & 134.998 & 2.15$^\dagger$ & 740466.857 & 74129.961 & 3.154 & 0.814--20.893 \\
GRB\,990821B & 17.000 & 47.724 & 2.15$^\dagger$ & 741017.649 & 74129.961 & 3.151 & 0.813--20.878 \\
GRB\,990820A & 23.000 & 69.023 & 2.15$^\dagger$ & 741060.523 & 74129.961 & 3.151 & 0.813--20.876 \\
GRB\,990803A & 32.400 & 41.613 & 2.15$^\dagger$ & 742555.048 & 74129.961 & 3.145 & 0.811--20.834 \\
GRB\,990726A & 46.000 & 110.354 & 2.15$^\dagger$ & 743293.039 & 74129.961 & 3.142 & 0.811--20.814 \\
GRB\,990720B & 35.000 & 88.409 & 2.15$^\dagger$ & 743791.622 & 74129.961 & 3.139 & 0.810--20.800 \\
GRB\,990713A & 44.000 & 84.097 & 2.15$^\dagger$ & 744391.732 & 74129.961 & 3.137 & 0.809--20.783 \\
GRB\,990711A & 59.000 & 132.316 & 2.15$^\dagger$ & 744573.351 & 74129.961 & 3.136 & 0.809--20.778 \\
GRB\,990701A & 31.000 & 84.737 & 2.15$^\dagger$ & 745423.966 & 74129.961 & 3.133 & 0.808--20.754 \\
GRB\,990622C* & 18.710 & 51.805 & 1.88$^\ddagger$ & 746183.503 & 71725.754 & 2.768 & 0.706--18.612 \\
GRB\,990622A & 34.000 & 38.210 & 2.15$^\dagger$ & 746203.586 & 74129.961 & 3.129 & 0.808--20.732 \\
GRB\,990620A & 15.000 & 38.808 & 2.15$^\dagger$ & 746332.241 & 74129.961 & 3.129 & 0.807--20.729 \\
GRB\,990606A & 43.000 & 74.745 & 2.15$^\dagger$ & 747619.484 & 74129.961 & 3.123 & 0.806--20.693 \\
GRB\,990603C & 21.000 & 45.942 & 2.15$^\dagger$ & 747815.315 & 74129.961 & 3.123 & 0.806--20.688 \\
GRB\,990603B & 17.100 & 45.644 & 2.15$^\dagger$ & 747816.327 & 74129.961 & 3.123 & 0.806--20.688 \\
GRB\,990506B & 35.000 & 91.924 & 2.15$^\dagger$ & 750232.161 & 74129.961 & 3.112 & 0.803--20.621 \\
GRB\,990319A & 47.000 & 55.086 & 2.15$^\dagger$ & 754403.657 & 74129.961 & 3.095 & 0.799--20.507 \\
GRB\,990312A & 53.000 & 132.455 & 2.15$^\dagger$ & 755036.674 & 74129.961 & 3.093 & 0.798--20.490 \\
GRB\,990310A & 43.000 & 114.963 & 2.15$^\dagger$ & 755196.125 & 74129.961 & 3.092 & 0.798--20.486 \\
GRB\,990130A & 31.000 & 77.868 & 2.15$^\dagger$ & 758548.191 & 74129.961 & 3.078 & 0.794--20.395 \\
GRB\,990129A* & 12.210 & 13.011 & 0.50$^\dagger$ & 758669.714 & 36584.175 & 0.723 & 0.175--5.439 \\
GRB\,990122A & 32.000 & 48.867 & 2.15$^\dagger$ & 759258.000 & 74129.961 & 3.075 & 0.794--20.376 \\
GRB\,981230A* & 18.350 & 53.231 & 0.50$^\dagger$ & 761194.019 & 36584.175 & 0.721 & 0.175--5.421 \\
GRB\,981219B & 44.000 & 124.618 & 0.50$^\dagger$ & 762155.479 & 36584.175 & 0.720 & 0.175--5.414 \\
GRB\,981129A* & 4.550 & 9.308 & 1.88$^\ddagger$ & 763896.753 & 71725.754 & 2.704 & 0.689--18.181 \\
GRB\,981126C* & 3.810 & 11.073 & 1.88$^\ddagger$ & 764180.519 & 71725.754 & 2.703 & 0.689--18.174 \\
GRB\,981126B & 34.000 & 38.221 & 2.15$^\dagger$ & 764143.413 & 74129.961 & 3.056 & 0.789--20.246 \\
GRB\,981126A & 39.000 & 67.394 & 2.15$^\dagger$ & 764150.134 & 74129.961 & 3.056 & 0.789--20.246 \\
GRB\,981104A & 49.000 & 105.132 & 2.15$^\dagger$ & 766051.175 & 74129.961 & 3.048 & 0.787--20.195 \\
GRB\,981101A & 25.800 & 75.304 & 2.15$^\dagger$ & 766353.441 & 74129.961 & 3.047 & 0.786--20.187 \\
GRB\,981030A & 39.000 & 89.483 & 2.15$^\dagger$ & 766540.612 & 74129.961 & 3.046 & 0.786--20.182 \\
GRB\,981017A & 41.000 & 112.397 & 0.50$^\dagger$ & 767666.327 & 36584.175 & 0.715 & 0.173--5.375 \\
GRB\,980916A & 15.500 & 36.607 & 2.15$^\dagger$ & 770273.689 & 74129.961 & 3.032 & 0.782--20.085 \\
GRB\,980907A & 13.700 & 16.153 & 2.15$^\dagger$ & 771084.227 & 74129.961 & 3.028 & 0.781--20.063 \\
GRB\,980728A & 33.000 & 48.457 & 2.15$^\dagger$ & 774635.299 & 74129.961 & 3.014 & 0.778--19.971 \\
GRB\,980720A & 36.000 & 58.440 & 2.15$^\dagger$ & 775278.047 & 74129.961 & 3.012 & 0.777--19.955 \\
GRB\,980718A & 30.000 & 48.383 & 2.15$^\dagger$ & 775526.776 & 74129.961 & 3.011 & 0.777--19.949 \\
GRB\,980714A & 46.000 & 122.117 & 2.15$^\dagger$ & 775805.542 & 74129.961 & 3.010 & 0.777--19.941 \\
GRB\,980701A & 37.000 & 61.885 & 2.15$^\dagger$ & 776932.665 & 74129.961 & 3.006 & 0.776--19.912 \\
GRB\,980622A & 44.000 & 130.562 & 2.15$^\dagger$ & 777704.115 & 74129.961 & 3.003 & 0.775--19.893 \\
GRB\,980615B & 20.000 & 57.213 & 2.15$^\dagger$ & 778345.477 & 74129.961 & 3.000 & 0.774--19.876 \\
GRB\,980419A & 39.000 & 88.021 & 2.15$^\dagger$ & 783230.888 & 74129.961 & 2.981 & 0.769--19.752 \\
GRB\,980416B* & 7.700 & 7.982 & 0.50$^\dagger$ & 783544.119 & 36584.175 & 0.700 & 0.170--5.267 \\
GRB\,980416A & 47.000 & 87.994 & 2.15$^\dagger$ & 783515.057 & 74129.961 & 2.980 & 0.769--19.745 \\
GRB\,980407B* & 7.920 & 13.906 & 0.50$^\dagger$ & 784277.997 & 36584.175 & 0.700 & 0.170--5.262 \\
GRB\,980407A & 39.000 & 45.528 & 2.15$^\dagger$ & 784275.914 & 74129.961 & 2.977 & 0.768--19.726 \\
GRB\,980406B & 35.000 & 68.811 & 2.15$^\dagger$ & 784368.169 & 74129.961 & 2.977 & 0.768--19.724 \\
GRB\,980324A & 25.000 & 45.800 & 2.15$^\dagger$ & 785512.329 & 74129.961 & 2.973 & 0.767--19.695 \\
GRB\,980320A & 37.000 & 38.662 & 2.15$^\dagger$ & 785840.065 & 74129.961 & 2.971 & 0.767--19.687 \\
GRB\,980228A & 33.000 & 92.892 & 2.15$^\dagger$ & 787543.515 & 74129.961 & 2.965 & 0.765--19.644 \\
GRB\,980224A & 42.000 & 94.008 & 2.15$^\dagger$ & 787897.359 & 74129.961 & 2.964 & 0.765--19.635 \\
GRB\,980218B* & 2.040 & 4.223 & 0.50$^\dagger$ & 788436.248 & 36584.175 & 0.696 & 0.169--5.234 \\
GRB\,980125A* & 3.330 & 9.709 & 2.15$^\dagger$ & 790527.348 & 74129.961 & 2.954 & 0.762--19.570 \\
GRB\,980116A & 52.000 & 142.411 & 1.88$^\ddagger$ & 791288.118 & 71725.754 & 2.611 & 0.665--17.551 \\
GRB\,971228A & 41.000 & 46.190 & 2.15$^\dagger$ & 792958.873 & 74129.961 & 2.945 & 0.760--19.510 \\
GRB\,971219A & 44.000 & 118.866 & 0.50$^\dagger$ & 793722.936 & 36584.175 & 0.691 & 0.168--5.199 \\
GRB\,971208A & 20.000 & 43.797 & 2.15$^\dagger$ & 794694.582 & 74129.961 & 2.938 & 0.758--19.467 \\
GRB\,971206C & 26.000 & 39.192 & 0.50$^\dagger$ & 794805.550 & 36584.175 & 0.690 & 0.167--5.192 \\
GRB\,971118A & 46.000 & 59.150 & 2.15$^\dagger$ & 796365.223 & 74129.961 & 2.932 & 0.757--19.427 \\
GRB\,971114A & 31.000 & 52.370 & 2.15$^\dagger$ & 796740.950 & 74129.961 & 2.931 & 0.756--19.417 \\
GRB\,971027B & 54.000 & 114.025 & 2.15$^\dagger$ & 798308.669 & 74129.961 & 2.925 & 0.755--19.379 \\
GRB\,971024B & 36.000 & 90.898 & 2.15$^\dagger$ & 798551.412 & 74129.961 & 2.924 & 0.755--19.373 \\
GRB\,971022B & 37.000 & 63.747 & 2.15$^\dagger$ & 798701.791 & 74129.961 & 2.924 & 0.754--19.370 \\
GRB\,971022A & 34.000 & 80.058 & 2.15$^\dagger$ & 798729.575 & 74129.961 & 2.924 & 0.754--19.369 \\
GRB\,971020A & 46.000 & 63.493 & 0.50$^\dagger$ & 798930.785 & 36584.175 & 0.687 & 0.167--5.165 \\
GRB\,971019A & 24.000 & 60.152 & 2.15$^\dagger$ & 798978.061 & 74129.961 & 2.923 & 0.754--19.363 \\
GRB\,970930A & 29.400 & 53.481 & 2.15$^\dagger$ & 800616.857 & 74129.961 & 2.917 & 0.753--19.323 \\
GRB\,970924A & 38.000 & 46.203 & 2.15$^\dagger$ & 801182.278 & 74129.961 & 2.915 & 0.752--19.310 \\
GRB\,970825B & 2.700 & 5.528 & 2.15$^\dagger$ & 803704.454 & 74129.961 & 2.905 & 0.750--19.249 \\
GRB\,970821A & 27.000 & 32.034 & 2.15$^\dagger$ & 804089.850 & 74129.961 & 2.904 & 0.749--19.240 \\
GRB\,970802B* & 4.300 & 9.605 & 2.15$^\dagger$ & 805710.113 & 74129.961 & 2.898 & 0.748--19.201 \\
GRB\,970518A & 14.000 & 30.127 & 2.15$^\dagger$ & 812311.483 & 74129.961 & 2.875 & 0.742--19.045 \\
GRB\,970506A & 34.000 & 63.469 & 0.50$^\dagger$ & 813317.707 & 36584.175 & 0.675 & 0.164--5.074 \\
GRB\,970424A & 47.000 & 91.095 & 2.15$^\dagger$ & 814375.230 & 74129.961 & 2.867 & 0.740--18.997 \\
GRB\,970326B & 31.000 & 32.201 & 2.15$^\dagger$ & 816851.618 & 74129.961 & 2.859 & 0.738--18.939 \\
GRB\,970314A & 40.000 & 40.386 & 1.88$^\ddagger$ & 817938.943 & 71725.754 & 2.525 & 0.644--16.979 \\
GRB\,970221A & 38.400 & 87.203 & 2.15$^\dagger$ & 819754.056 & 74129.961 & 2.849 & 0.735--18.872 \\
GRB\,970128A & 40.000 & 67.009 & 2.15$^\dagger$ & 821756.506 & 74129.961 & 2.842 & 0.733--18.826 \\
GRB\,961218B* & 12.690 & 14.325 & 2.15$^\dagger$ & 825313.285 & 74129.961 & 2.829 & 0.730--18.745 \\
GRB\,961125B & 38.000 & 108.634 & 2.15$^\dagger$ & 827342.949 & 74129.961 & 2.822 & 0.728--18.699 \\
GRB\,961020A* & 9.300 & 26.134 & 0.50$^\dagger$ & 830471.116 & 36584.175 & 0.661 & 0.160--4.969 \\
GRB\,961017A* & 8.120 & 20.955 & 0.50$^\dagger$ & 830718.514 & 36584.175 & 0.661 & 0.160--4.967 \\
GRB\,961011A & 47.000 & 110.639 & 2.15$^\dagger$ & 831188.680 & 74129.961 & 2.809 & 0.725--18.613 \\
GRB\,960916C & 60.000 & 104.801 & 2.15$^\dagger$ & 833388.344 & 74129.961 & 2.802 & 0.723--18.564 \\
GRB\,960801B & 50.000 & 135.285 & 2.15$^\dagger$ & 837336.473 & 74129.961 & 2.789 & 0.720--18.476 \\
GRB\,960703A & 36.000 & 47.171 & 2.15$^\dagger$ & 839871.426 & 74129.961 & 2.780 & 0.717--18.420 \\
GRB\,960513A* & 14.760 & 20.544 & 0.50$^\dagger$ & 844225.155 & 36584.175 & 0.650 & 0.158--4.888 \\
GRB\,951124B* & 15.480 & 40.818 & 0.50$^\dagger$ & 859040.883 & 36584.175 & 0.639 & 0.155--4.804 \\
GRB\,951030B* & 12.340 & 28.850 & 1.88$^\ddagger$ & 861208.637 & 71725.754 & 2.399 & 0.611--16.126 \\
GRB\,950313A* & 1.690 & 5.050 & 1.88$^\ddagger$ & 881170.762 & 71725.754 & 2.344 & 0.597--15.761 \\
GRB\,950102B* & 4.460 & 6.012 & 0.50$^\dagger$ & 887218.828 & 36584.175 & 0.619 & 0.150--4.651 \\
GRB\,941123A* & 12.180 & 35.579 & 0.50$^\dagger$ & 890634.738 & 36584.175 & 0.616 & 0.149--4.633 \\
GRB\,940917A* & 3.450 & 9.883 & 2.15$^\dagger$ & 896428.852 & 74129.961 & 2.605 & 0.672--17.258 \\
GRB\,940825A* & 12.590 & 21.575 & 0.50$^\dagger$ & 898473.967 & 36584.175 & 0.611 & 0.148--4.593 \\
GRB\,940114B* & 4.580 & 7.631 & 1.88$^\ddagger$ & 917668.115 & 71725.754 & 2.251 & 0.574--15.134 \\
GRB\,940108A* & 10.230 & 24.450 & 0.50$^\dagger$ & 918240.120 & 36584.175 & 0.598 & 0.145--4.494 \\
GRB\,930903A* & 4.200 & 6.139 & 2.15$^\dagger$ & 929235.451 & 74129.961 & 2.513 & 0.648--16.649 \\
GRB\,930601A* & 7.860 & 11.070 & 1.88$^\ddagger$ & 937335.962 & 71725.754 & 2.204 & 0.562--14.817 \\
GRB\,930528A* & 6.030 & 16.745 & 0.50$^\dagger$ & 937659.463 & 36584.175 & 0.585 & 0.142--4.401 \\
GRB\,930421A* & 14.590 & 26.583 & 0.50$^\dagger$ & 940884.597 & 36584.175 & 0.583 & 0.141--4.386 \\
GRB\,930331C* & 11.490 & 25.920 & 1.88$^\ddagger$ & 942657.006 & 71725.754 & 2.191 & 0.558--14.733 \\
GRB\,930318A* & 5.590 & 16.164 & 2.15$^\dagger$ & 943812.457 & 74129.961 & 2.474 & 0.638--16.392 \\
GRB\,930131B* & 5.250 & 15.320 & 2.15$^\dagger$ & 947736.348 & 74129.961 & 2.464 & 0.636--16.324 \\
GRB\,930114B* & 5.870 & 13.806 & 1.88$^\ddagger$ & 949265.403 & 71725.754 & 2.176 & 0.555--14.630 \\
GRB\,921222A* & 4.850 & 11.513 & 2.15$^\dagger$ & 951194.645 & 74129.961 & 2.455 & 0.633--16.264 \\
GRB\,921002A* & 25.950 & 40.209 & 1.88$^\ddagger$ & 958225.655 & 71725.754 & 2.156 & 0.549--14.494 \\
GRB\,920123A* & 8.120 & 24.000 & 1.88$^\ddagger$ & 980084.524 & 71725.754 & 2.108 & 0.537--14.170 \\
GRB\,910912A* & 6.000 & 15.315 & 0.50$^\dagger$ & 991596.047 & 36584.175 & 0.553 & 0.134--4.162 \\
GRB\,910823A* & 13.080 & 19.487 & 2.15$^\dagger$ & 993269.741 & 74129.961 & 2.351 & 0.607--15.575 \\
GRB\,910629A* & 26.560 & 54.148 & 2.15$^\dagger$ & 998078.476 & 74129.961 & 2.340 & 0.604--15.500 \\
GRB\,910502A* & 13.330 & 26.015 & 0.50$^\dagger$ & 1003067.951 & 36584.175 & 0.547 & 0.133--4.114 \\
GRB\,910427A* & 9.070 & 17.001 & 2.15$^\dagger$ & 1003508.258 & 74129.961 & 2.327 & 0.600--15.417 \\
\bottomrule
\addlinespace[6pt]
\multicolumn{8}{c}{\parbox{0.8\textwidth}{\footnotesize * denotes GRBs without GCN-style names (auto-generated by GRBweb). $^a$ marks GRBs without final recognized position error, the errors given are estimates based on detectors~\cite{von_Kienlin_2020}. $^\ddagger$ shows estimated average redshift values $z=1.88$ for GRBs lacking redshift/duration data~\cite{10.3389/fspas.2023.1124317}, and $^\dagger$ denotes average redshift estimates ($z=2.15$ for long bursts; $z=0.5$ for short bursts).}}
\end{longtable}

\emph{\textbf{Acknowledgements}.---}%
This work is supported by National Natural Science Foundation of China under grant No.~12335006. 
This work is also supported by High-performance Computing Platform of Peking University.

%



\bibliographystyle{aasjournal}

\bibliography{scibib}{}


\end{document}